\documentclass[journal]{IEEEtran}
\usepackage[utf8]{inputenc}
\usepackage{amsmath}
\usepackage{amssymb}
\usepackage{amsbsy}
\usepackage{graphicx}
\usepackage{url}
\usepackage{textcomp}
\usepackage{xcolor}
\usepackage{hyperref}
\usepackage{cleveref}

\newcommand{\hadamard}{\odot}

\newcommand{\herm}[1]{{#1}^{H}}

\begin{document}

\title{\huge Practical Channel Estimation for Pinching-Antenna Systems: Serial vs. Parallel and Downlink vs. Uplink?
	\thanks{
	Jian Xiao is with the Department of Electronics and Information Engineering, College of Physical Science and Technology, Central China Normal University, Wuhan 430079, China (e-mail: jianx@mails.ccnu.edu.cn). 
	}}
\author{
    \IEEEauthorblockN{Jian Xiao}
}

\maketitle

\begin{abstract}
The practical channel estimation for pinching-antenna networks is investigated, in which an electromagnetic-compliant in-waveguide transmission model is exhibited, incorporating bidirectional power splitting, cumulative power leakage, and waveguide attenuation. Based on this model, the paper investigates two antenna activation protocols for channel estimation: a serial protocol based on one-by-one antenna activation and a parallel protocol utilizing a binary S-Matrix activation. The serial protocol is characterized by its superior numerical stability but a lack of array gain, whereas the parallel protocol theoretically offers array gain but suffers from severe performance degradation due to structural crosstalk from the non-orthogonal S-Matrix and ill-conditioning from cumulative leakage. Furthermore, the paper analyzes the fundamental commonalities and asymmetries between uplink and downlink channel estimation in pinching-antenna systems. Numerical results demonstrate that 1) in an ideal lossless model, the parallel protocol is superior to the serial protocol due to the array gain from simultaneous energy collection in uplink transmission; 2) in a practical model with physical losses, the serial protocol outperforms the parallel protocol, as the performance of the parallel protocol is degraded by the numerical instability from cumulative leakage, which outweighs the benefit of array gain; 3) For downlink channel estimation, the serial protocol is more suitable because its strategy of concentrating the entire power budget on one measurement, while the parallel protocol is more suitable for the uplink as it can make full use of array gain.
\end{abstract}


\section{Introduction}
The relentless pursuit of higher data rates in future sixth-generation (6G) systems has spurred innovation in antenna technologies. Pinching antennas (PAs) have emerged as a promising candidate \cite{Ding2024}, offering a flexible and scalable method for deploying communication areas, particularly in high-frequency bands. Pinching-antenna networks (PANs) utilize low-loss dielectric waveguides as the signal bus, from which radio waves can be controllably radiated at arbitrary locations by applying a pinching mechanism. Despite its architectural flexibility, the performance of PANs is critically dependent on the availability of accurate channel state information. 
The channel estimation for PANs can be logically decomposed into a two-step process. The first step involves estimating the channel for the currently activated antennas by using pilot signals. The second step, which is unique to this antenna architecture, involves extrapolating these measurements to estimate the channels for all potential antenna positions along the entire waveguide. This allows for dynamic and flexible beamforming without needing to physically activate every possible antenna position during the initial estimation phase. However, the physical behavior of the shared waveguide is complex and highly dependent on the operational mode \cite{pozar2021microwave}, leading to significant challenges.

\begin{figure}[t]
	\centerline{\includegraphics[width=2.7in]{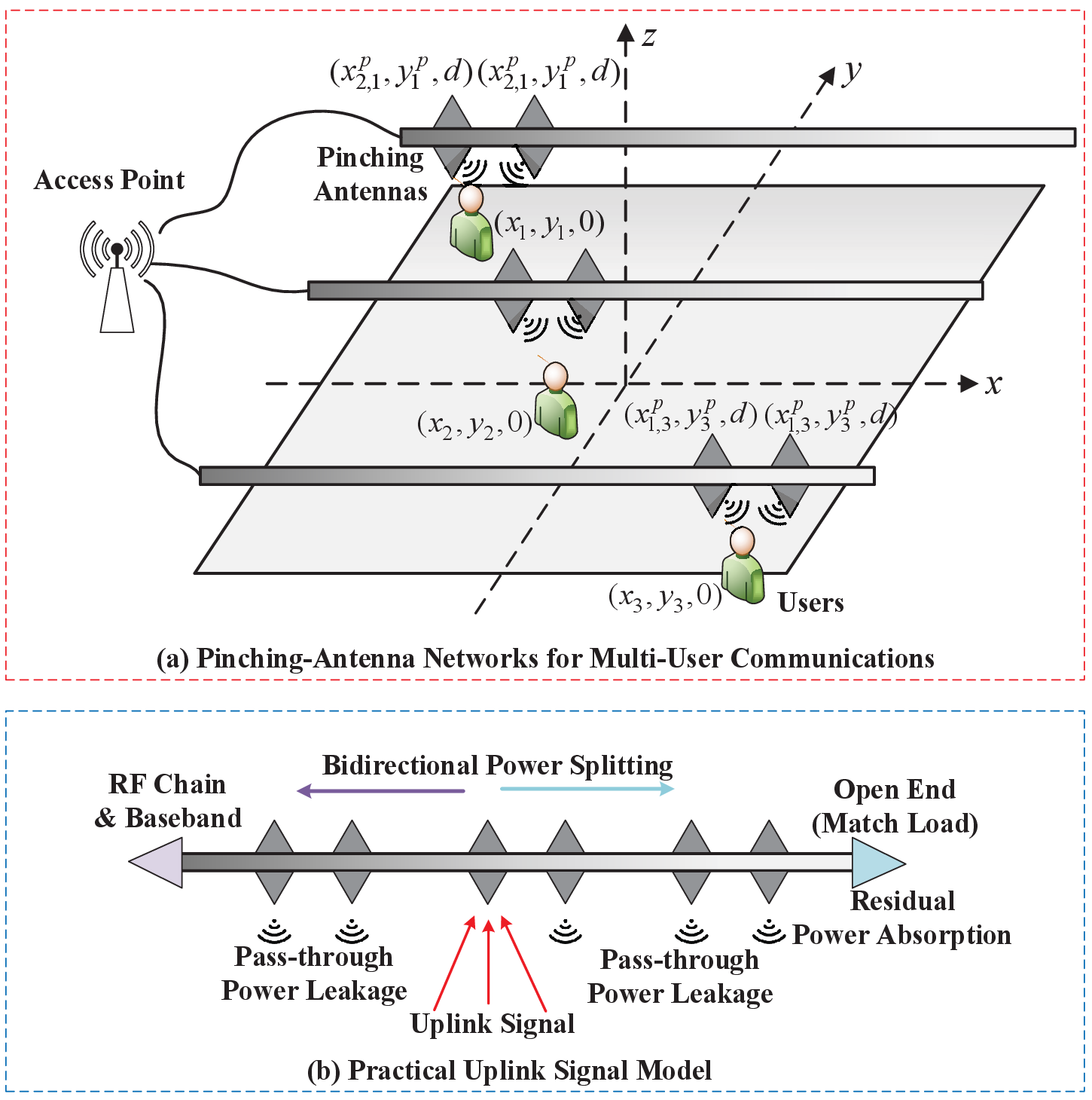}}
	\caption{{Illustration of pinching-antenna networks.}}
	\label{fig1_2}
\end{figure}

As demonstrated in practical pinching-antenna experiments \cite{Fukuda2022}, a single and isolated pinching point can be created on demand, with the rest of the waveguide acting as a simple transmission line. Conversely, when multiple antennas are active, the system may behave as a coupled leaky-wave structure, where signals propagating along the waveguide suffer from cumulative leakage as they pass other antenna elements. Although preliminary works have investigated the uplink channel estimation in PANs \cite{11018390, liu2025pinching}, they often rely on an idealized in-waveguide transmission model. Specifically, considering a waveguide equipped with $N$ PAs, the idealized in-waveguide channel is given by
\begin{equation}
\begin{split}\label{waveguide}
\mathbf{g} =\begin{bmatrix}
\alpha_{1} e^{-j \tfrac{2\pi}{\lambda_g} \left|\psi_{0}^{\rm{P}} - \psi_{1}^{\rm{P}} \right|}, 
\ldots, 
\alpha_{N} e^{-j \tfrac{2\pi}{\lambda_g} \left|\psi_{0}^{\rm{P}} - \psi_{N}^{\rm{P}} \right|}\end{bmatrix}^\top,
\end{split}
\end{equation}
where \( \psi _{n}^{\rm{P}}=(x_{n}^{\rm{P}},y_n^{\rm{P}},d), 1\le n \le N\) denotes the position of PA $n$, \(\psi_{0}^{\rm{P}}\) denotes the position of the feed point of the $m$-th waveguide, \(\lambda_g\) is the guided wavelength, and $\alpha_{n}$ is the power exchanged factor between the waveguide and PA $n$.

The above ideal model assumes that uplink signals received by different PAs on the same waveguide propagate to the feed point independently. However, in a practical physical scenario, the shared waveguide acts as a common bus, leading to coupling and leakage effects. As illustrated in Fig.~\ref{fig1_2}, a signal coupled into the waveguide by one PA will be partially leaked out by other PAs it passes on its way to the feed point according to the coupled-mode theory and the principle of reciprocity \cite{collin1990field,104225}. Moreover, when a signal is coupled from a PA into the waveguide, the induced wave propagates bidirectionally. By incorporating both the cumulative leakage along the propagation path and the initial bidirectional power split at the coupling point, the $n$-th element of the practical in-waveguide channel can be expressed as
\begin{equation}
\begin{split}\label{comprehensive_practical_waveguide}
[\mathbf{g}']_n &= \underbrace{\sqrt{\gamma}}_{\text{Bidirectional Split}} \cdot \underbrace{\left( \prod_{i=1}^{n-1} \beta_{i} \right)}_{\text{Cumulative Pass-through}} \cdot \underbrace{\alpha_{n}}_{\text{Coupling-in}} \\ 
& \cdot \underbrace{e^{-\varepsilon |\psi_{0}^{\rm{P}}-\psi_{n}^{\rm{P}}|}}_{\text{Waveguide Attenuation}} \cdot \underbrace{e^{-j\frac{2\pi}{\lambda_g}|\psi_{0}^{\rm{P}}-\psi_{n}^{\rm{P}}|}}_{\text{Propagation Phase}}\\
&=\sqrt{\gamma} \alpha_{n} \left( \prod_{i=1}^{n-1} \beta_{i} \right) e^{-(\varepsilon + j\frac{2\pi}{\lambda_g})|\psi_{0}^{\rm{P}}-\psi_{n}^{\rm{P}}|},
\end{split}
\end{equation}
where $\gamma \in [0, 1]$ denotes a bidirectional power division factor, representing the fraction of coupled power that propagates towards the access point (AP), $\varepsilon$ denotes the attenuation factor in-waveguide propagation, and $\beta_{i},i=\{1,\ldots,n-1\}$ is the pass-through factor for the PA $n$. According to the coupling reciprocity of PAs, the power exchange principle follows $\beta_{i}^2 +  \alpha_{i}^2 = 1$. 

This refined model reveals that effective uplink communication requires overcoming both the initial coupling loss and the cumulative propagation leakage. Ignoring this model mismatch between the ideal assumption and physical reality can lead to flawed channel estimation, which in turn degrades the performance of higher-level tasks such as beamforming and data detection. This paper bridges the gap by developing a comprehensive theoretical framework for channel estimation based on the practical transmission model in PANs. We aim to answer the following fundamental questions:
\begin{itemize}
    \item \textbf{Q1:} \textit{What are the performance implications of the practical transmission model compared to the ideal model for PANs?}
    \item \textbf{Q2:} \textit{Under this practical model, how to design the PA activation strategy for channel estimation? }
    \item \textbf{Q3:} \textit{Considering estimation accuracy and overhead, should uplink or downlink channel estimation be employed for PANs?}
\end{itemize}

To address the aforementioned questions, we first establish a electromagnetic-compliant in-waveguide transmission model, which incorporates key physical factors according to the classic transmission line model and coupled-mode theory \cite{pozar2021microwave}. Furthermore, we propose two PA activation protocols for channel estimation, i.e., serial (one-by-one) and parallel (simultaneous) activation strategies. We provide a rigorous  performance analysis demonstrating that the choice between serial and parallel estimation hinges on a critical trade-off between array gain and the numerical stability of the estimator, which is dictated by the leakage characteristics of the PA hardware. Finally, we prove that a profound asymmetry exists between uplink and downlink estimation. Uplink is an energy-collection problem where the aforementioned stability trade-off is central. In contrast, downlink is a power-budget-limited problem where serial activation is unconditionally superior due to its ability to concentrate the total transmit power on a single measurement. Our findings, validated by numerical results, conclude that under the constraint of physically realizable binary switching and the practical in-waveguide transmission model, the robust serial protocol consistently outperforms the S-Matrix-based parallel protocol in both uplink and downlink scenarios.

\section{System Models}
For clarity of derivation and without loss of generality, we first analyze the case of a single waveguide. The results can be directly extended to the multi-waveguide case as the estimation on each waveguide can be considered an independent process.  We consider an AP is equipped with a waveguide to serve $K$ single-antenna user equipments (UEs). The waveguide contains $N$ controllable PAs. The waveguide is installed at a height \( d \). In a three-dimensional Cartesian system, the UEs are assumed to be randomly distributed within a rectangular region on the $x$-$y$ plane, with dimensions $D_x$ and $D_y$. The position of the $k$-th UE is represented by $\psi_k = (x_k, y_k, 0)$.

\subsection{Wireless Propagation Channel}
The wireless channel vector $\mathbf{h}_{k} \in \mathbb{C}^{N \times 1}$ from the $k$-th UE to the $N$ PAs is composed of Line-of-Sight (LoS) and Non-Line-of-Sight (NLoS) components:
\begin{equation}
\mathbf{h}_{k} = \mathbf{h}_{k}^{\text{LoS}} + \mathbf{h}_{k}^{\text{NLoS}},
\label{eq:wireless_channel_total}
\end{equation}
The LoS component is modeled considering a visibility region vector $\boldsymbol{\Upsilon}_{k} = [\upsilon_{k,1}, \dots, \upsilon_{k,N}]^T \in \{0,1\}^N$ (where $\upsilon_{k,n}=1$ if the $n$-th PA is in the LoS of the $k$-th UE, and 0 otherwise):
\begin{equation}
\mathbf{h}_{k}^{\text{LoS}} = \boldsymbol{\Upsilon}_{k} \hadamard \left[ \frac{\sqrt{\beta_0}e^{-j\frac{2\pi}{\lambda}|\psi_{k}-\psi_{1}^{P}|}}{|\psi_{k}-\psi_{1}^{P}|}, \dots, \frac{\sqrt{\beta_0}e^{-j\frac{2\pi}{\lambda}|\psi_{k}-\psi_{N}^{P}|}}{|\psi_{k}-\psi_{N}^{P}|} \right]^T,
\label{eq:wireless_channel_los}
\end{equation}
where $\beta_0$ is the free-space path loss constant and $\lambda$ is the carrier wavelength. The NLoS component, comprising $S$ scatterers, is given by
\begin{equation}
\mathbf{h}_{k}^{\text{NLoS}} = \sqrt{\frac{\beta_0}{S}} \sum_{s=1}^{S} \xi_{k,s} \mathbf{a}_{k,s},
\label{eq:wireless_channel_nlos}
\end{equation}
where $\xi_{k,s} \sim \mathcal{CN}(0, \sigma_s^2)$ is the complex gain of the $s$-th scatterer path. The array response vector $\mathbf{a}_{k,s} \in \mathbb{C}^{N \times 1}$ under the spherical wave model is given by
\begin{equation}
\mathbf{a}_{k,s} = \left[ \frac{e^{-j\frac{2\pi}{\lambda}d_{s,m,1}}}{d_{s,1}}, \dots, \frac{e^{-j\frac{2\pi}{\lambda}d_{s,N}}}{d_{s,N}} \right]^T,
\label{eq:array_response}
\end{equation}
where $d_{s,n}$ denotes the distance from the $s$-th scatterer to the $n$-th PA.

\subsection{Uplink In-Waveguide Channel Models}
We define two distinct models for the uplink in-waveguide transfer vector $\mathbf{g} \in \mathbb{C}^{N \times 1}$, which describes the signal propagation from the PAs to the AP.

\subsubsection{Controllable Radiation Model for Single PA Activation}
This model assumes that when only a single PA is activated (serial mode), inter-antenna coupling is negligible, and the waveguide acts as a near-ideal transmission line. The signal captured by the $n$-th PA suffers two main losses: an initial bidirectional power splitting loss $\gamma$ and intrinsic waveguide attenuation that is proportional to in-waveguide propagation distance. 
The $n$-th element of of $[\mathbf{g}_{\text{serial}}]$ is given by
\begin{equation}
[\mathbf{g}_{\text{serial}}]_n = \sqrt{\gamma} {\alpha_{n}} e^{-\varepsilon d_n} e^{-j\frac{2\pi}{\lambda_g}d_n}.
\label{eq:serial_transfer_complex}
\end{equation}
where $d_n$ is the distance from the $n$-th PA to the AP, $\varepsilon$ is the intrinsic attenuation constant of the waveguide, and $\lambda_g$ is the guided wavelength (dependent on waveguide geometry and carrier frequency). Suppose the waveguide mode and the PA mode has the perfect phase matching, the power exchanged factor between the waveguide and the $n$-th PA $\alpha_{n}$ can be expressed as $\alpha_{n}= \sin(\varpi_{n} L_{n})$ \cite{Wang2025}, where $\varpi_{n}$ and $L_{n}$ denote the coupling strength and length between the waveguide and the $n$-th PA.

\subsubsection{Cumulative Leakage Model for Multiple PA Activation}
This model assumes that when multiple PAs are active simultaneously (parallel mode), electromagnetic coupling between PAs becomes significant. The signal from the $n$-th PA must pass through PAs $1$ to $n-1$ to reach the AP, and each intermediate PA causes a fraction of the signal to leak out . Let $\beta_i = \sqrt{1- \alpha_{i}^2} = \cos(\kappa_{i} L_{i})$ be the signal pass-through coefficient for the $i$-th PA, i.e., a fraction $\beta_i$ of the signal passes through, and $1-\beta_i$ leaks out. The the $n$-th element of $\mathbf{g}_{\text{parallel}}$ is given by
\begin{equation}
[\mathbf{g}_{\text{parallel}}]_n = \sqrt{\gamma} {\alpha_{n}} e^{-\varepsilon d_n} e^{-j\frac{2\pi}{\lambda_g}d_n} \left( \prod_{i=1}^{n-1} \beta_i \right).
\label{eq:parallel_transfer_complex}
\end{equation}
The product term $\prod_{i=1}^{n-1} \beta_i$ captures the cumulative leakage effect: PAs further from the AP experience more significant attenuation due to multiple pass-through events.

\subsection{Downlink In-Waveguide Channel Model}
For the downlink, the signal originates at the AP, propagates along the waveguide, and is radiated by active PAs to UEs. This process involves two key losses: an initial feed coupling loss at the AP and cumulative radiation loss . Let $\eta \in (0,1)$ be the feed coupling efficiency at the AP, representing the fraction of BS signal coupled into the waveguide. The transfer vector $\mathbf{g}_{\text{downlink}} \in \mathbb{C}^{N \times 1}$ describes the signal arriving at the $n$-th PA for radiation. 
The downlink in-waveguide channel is given by
\begin{equation}
[\mathbf{g}_{\text{downlink}}]_n = \sqrt{\eta} \alpha_n e^{-\varepsilon d_n} e^{-j\frac{2\pi}{\lambda_g}d_n} \left( \prod_{i=1}^{n-1} \sqrt{1-\alpha_i^2} \right).
\label{eq:downlink_transfer_complex}
\end{equation}
This model is structurally symmetric to the uplink cumulative leakage model: the cumulative product term $\prod \sqrt{(1-\alpha_i^2)}$ for downlink radiation loss mirrors the uplink cumulative leakage term $\prod \beta_i$.

\textbf{Remark 1:}
 The channel reciprocity between downlink and uplink links does not hold for the end-to-end pinching-antenna system. This conclusion stems from the fundamental differences in the governing equations for the uplink and downlink hardware responses. The uplink transfer vector $\mathbf{g}_\text{uplink}$ is characterized by a bidirectional splitting factor $\gamma$ and cumulative pass-through leakage coefficients $\beta_i$. In contrast, the downlink transfer vector $\mathbf{g}_\text{downlink}$ is determined by a distinct feed coupling efficiency $\eta$ and cumulative radiation loss coefficients $\alpha_i$. Since the physical parameters and thus the mathematical forms of $\mathbf{g}_\text{uplink}$ and $\mathbf{g}_\text{downlink}$ are not equivalent, the condition for hardware reciprocity is violated.

\section{Uplink Channel Estimation Protocols}

\subsection{Problem Formulation}
The ultimate goal of uplink channel estimation is to recover the wireless channel vector $\mathbf{h}_k \in \mathbb{C}^{N \times 1}$, which captures the environment-dependent propagation characteristics and is time-varying. In contrast, the in-waveguide transfer vectors $\mathbf{g}_{\text{serial}}$ and $\mathbf{g}_{\text{parallel}}$ are hardware-dependent, determined by waveguide geometry, PA spacing, and material properties, and can be pre-calibrated offline, thus treated as {known} prior to estimation. We assume that time-orthogonal pilot sequences are assigned to the $K$ users, which allows for interference-free channel estimation for each user individually. In this work, the subsequent analysis focuses on the estimation process for an arbitrary user $k$ and the initial channel estimation for the currently activated antennas. Therefore, for notational simplicity, the user index $k$ is dropped from all relevant terms. The derived performance metrics are applicable to any user in the system. We define the composite channel $\mathbf{b}$ for each operational mode. For serial estimation, we define
  \begin{equation}
  \mathbf{b}_{\text{serial}} = \mathbf{g}_{\text{serial}} \hadamard \mathbf{h}.
  \label{eq:serial_composite_channel}
  \end{equation}
 For parallel estimation, we have
  \begin{equation}
  \mathbf{b}_{\text{parallel}} = \mathbf{g}_{\text{parallel}} \hadamard \mathbf{h}.
  \label{eq:parallel_composite_channel}
  \end{equation}

To explicitly model the relationship between wireless channel $\mathbf{h}$ and the received signal, we first define diagonal in-waveguide channel matrices converting vector Hadamard products to matrix multiplication, i.e., 
$\mathbf{G}_{\text{serial}} = \text{diag}(\mathbf{g}_{\text{serial}}) \in \mathbb{C}^{N \times N}$, and $
\mathbf{G}_{\text{parallel}} = \text{diag}(\mathbf{g}_{\text{parallel}}) \in \mathbb{C}^{N \times N}$
The composite channels can be rewritten as matrix-vector products:
\begin{equation}
\mathbf{b}_{\text{serial}} = \mathbf{G}_{\text{serial}} \mathbf{h}, \quad \mathbf{b}_{\text{parallel}} = \mathbf{G}_{\text{parallel}} \mathbf{h}.
\label{eq:composite_channel_matrix_form}
\end{equation}

The received signal vector over $N$ pilot slots follows a linear model:
\begin{equation}
\mathbf{y} = \mathbf{W} \mathbf{b} + \mathbf{n},
\label{eq:received_signal_general}
\end{equation}
where $\mathbf{W} \in \mathbb{R}^{N \times N}$ is a binary activation matrix, i.e., 1 indicates a PA is active in a slot, 0 otherwise, $\mathbf{b}$ is the composite channel, i.e., either $\mathbf{b}_{\text{serial}}$ or $\mathbf{b}_{\text{parallel}}$, and $\mathbf{n} \sim \mathcal{CN}(0, \sigma_n^2 \mathbf{I}_N)$ is the additive white Gaussian noise (AWGN) vector with variance $\sigma_n^2$.

Substituting \cref{eq:composite_channel_matrix_form} into \cref{eq:received_signal_general}, we derive a wireless channel-centric signal model as
\begin{equation}
\mathbf{y} = \mathbf{A} \mathbf{h} + \mathbf{n},
\label{eq:received_signal_h_centric}
\end{equation}
where $\mathbf{A} \in \mathbb{C}^{N \times N}$ is the observation matrix, defined as
\begin{equation}
\mathbf{A} = \mathbf{W} \mathbf{G},
\label{eq:observation_matrix_def}
\end{equation}
where $\mathbf{G} =\mathbf{G}_{\text{serial}}$ for serial estimation or $\mathbf{G} =\mathbf{G}_{\text{parallel}}$ for parallel estimation.

\subsection{Serial Estimation Protocol}
In serial mode, only one PA is activated per slot, so the activation matrix is identity matrix, i.e., $\mathbf{W}_{\text{serial}} = \mathbf{I}_N$. Substituting into \cref{eq:observation_matrix_def}, the observation matrix simplifies to
\begin{equation}
\mathbf{A}_{\text{serial}} = \mathbf{I}_N \mathbf{G}_{\text{serial}} = \mathbf{G}_{\text{serial}}.
\label{eq:serial_observation_matrix}
\end{equation}
The received signal model becomes
\begin{equation}
\mathbf{y}_{\text{serial}} = \mathbf{G}_{\text{serial}} \mathbf{h} + \mathbf{n}.
\label{eq:serial_received_signal_h}
\end{equation}

The least squares (LS) estimator minimizes the squared error $\|\mathbf{y}_{\text{serial}} - \mathbf{G}_{\text{serial}} \mathbf{h}\|_2^2$. Since $|[g_{\text{serial}}]_n| > 0$ for all $n$, for a full-rank $\mathbf{G}_{\text{serial}}$, the LS estimator is given by
\begin{equation}
\hat{\mathbf{h}}_{\text{serial}}^{\text{LS}} = \left( \herm{\mathbf{G}_{\text{serial}}} \mathbf{G}_{\text{serial}} \right)^{-1} \herm{\mathbf{G}_{\text{serial}}} \mathbf{y}_{\text{serial}}.
\label{eq:serial_ls_estimator_h}
\end{equation}

\subsection{Parallel Estimation Protocol}
In parallel mode, multiple PAs are activated per slot to exploit array gain. We develop an S-Matrix activation scheme derived from Hadamard matrices to ensure orthogonality of measurements, enabling unique recovery of the composite channel. The activation matrix is constructed as:
\begin{equation}
\mathbf{W}_{\text{parallel}} = \mathbf{W}_{\text{S-Matrix}} = \frac{1}{2} \left( \mathbf{H}_N + \mathbf{J}_N \right),
\label{eq:parallel_activation_matrix}
\end{equation}
where $\mathbf{H}_N$ is the $N \times N$ Hadamard matrix and $\mathbf{J}_N$ is the $N \times N$ matrix of all ones. This matrix is invertible, ensuring the composite channel can be uniquely recovered.

The received signal model becomes
\begin{equation}
\mathbf{y}_{\text{parallel}} = \mathbf{W}_{\text{S-Matrix}} \mathbf{G}_{\text{parallel}} \mathbf{h} + \mathbf{n},
\label{eq:parallel_received_signal_h}
\end{equation}

For a full-rank $\mathbf{A}_{\text{parallel}}$, ensured by the invertibility of $\mathbf{W}_{\text{S-Matrix}}$ and $\mathbf{G}_{\text{parallel}}$, the LS estimator is given by
\begin{equation}
\hat{\mathbf{h}}_{\text{parallel}}^{\text{LS}} = \left( \herm{\mathbf{A}_{\text{parallel}}} \mathbf{A}_{\text{parallel}} \right)^{-1} \herm{\mathbf{A}_{\text{parallel}}} \mathbf{y}_{\text{parallel}}.
\label{eq:parallel_ls_estimator_h_general}
\end{equation}
Substituting $\mathbf{A}_{\text{parallel}} = \mathbf{W}_{\text{S-Matrix}} \mathbf{G}_{\text{parallel}}$, this expands to
\begin{equation}
\hat{\mathbf{h}}_{\text{parallel}}^{\text{LS}} = \mathbf{V}
\herm{\mathbf{G}_{\text{parallel}}} \herm{\mathbf{W}_{\text{S-Matrix}}} \mathbf{y}_{\text{parallel}},
\label{eq:parallel_ls_estimator_h_expanded}
\end{equation}
where $\mathbf{V} = \left( \herm{\mathbf{G}_{\text{parallel}}} \herm{\mathbf{W}_{\text{S-Matrix}}} \mathbf{W}_{\text{S-Matrix}} \mathbf{G}_{\text{parallel}} \right)^{-1}$.

\section{Performance Analysis of Serial and Parallel Channel Estimator}

\subsection{MSE of Uplink LS Estimators}
For the general LS estimator $\hat{\mathbf{h}} = \left( \herm{\mathbf{A}} \mathbf{A} \right)^{-1} \herm{\mathbf{A}} \mathbf{y}$, the mean squared error (MSE) is given by the trace of the error covariance matrix:
\begin{equation}
\text{MSE} = \text{Tr}\left( \mathbb{E}\left[ (\hat{\mathbf{h}} - \mathbf{h}) \herm{(\hat{\mathbf{h}} - \mathbf{h})} \right] \right) = \sigma_n^2 \text{Tr}\left( \left( \herm{\mathbf{A}} \mathbf{A} \right)^{-1} \right),
\label{eq:general_ls_mse}
\end{equation}
This formula directly links the MSE to the conditioning of the observation matrix $\mathbf{A}$, i.e., a smaller trace of $(\herm{\mathbf{A}} \mathbf{A})^{-1}$ indicates better estimation performance.

\subsubsection{MSE of Serial Estimator for $\mathbf{h}$}
From \cref{eq:serial_observation_matrix}, $\mathbf{A}_{\text{serial}} = \mathbf{G}_{\text{serial}}$. Substituting into \cref{eq:general_ls_mse}, the MSE becomes
\begin{equation}
\text{MSE}_{\text{serial}} = \sigma_n^2 \text{Tr}\left( \left( \herm{\mathbf{G}_{\text{serial}}} \mathbf{G}_{\text{serial}} \right)^{-1} \right),
\label{eq:serial_mse_h_general}
\end{equation}

Since $\mathbf{G}_{\text{serial}}$ is diagonal, $\herm{\mathbf{G}_{\text{serial}}} \mathbf{G}_{\text{serial}}$ is also diagonal with elements $|[g_{\text{serial}}]_n|^2 = \gamma e^{-2\alpha d_n}$. The inverse of a diagonal matrix is diagonal with elements equal to the inverse of the original diagonal elements. Thus, we have
\begin{equation}
\left( \herm{\mathbf{G}_{\text{serial}}} \mathbf{G}_{\text{serial}} \right)^{-1} = \\
\text{diag}\left( \frac{1}{|[\mathbf{g}_{\text{serial}}]_1|^2}, \dots, \frac{1}{|[\mathbf{g}_{\text{serial}}]_N|^2} \right).
\label{eq:serial_gram_inverse}
\end{equation}

Substituting \cref{eq:serial_gram_inverse} into \cref{eq:serial_mse_h_general}, the serial MSE simplifies to a sum of per-PA error terms:
\begin{equation}
\text{MSE}_{\text{serial}} = \sigma_n^2 \sum_{n=1}^{N} \frac{1}{|[\mathbf{g}_{\text{serial}}]_n|^2} = \sigma_n^2 \sum_{n=1}^{N} \frac{1}{\gamma \alpha_{n}^2e^{-2\varepsilon d_n}}.
\label{eq:serial_mse_h_expanded}
\end{equation}

For serial channel estimation, each term in the sum is $\sigma_n^2 / |[\mathbf{g}_{\text{serial}}]_n|^2$, which is the inverse of the measurement SNR for the $n$-th PA. Since serial estimation uses only one PA per slot, each measurement has low SNR, leading to a large cumulative MSE.

\subsubsection{MSE of Parallel Estimator for $\mathbf{h}$}
Substituting $\mathbf{A}_{\text{parallel}} = \mathbf{W}_{\text{S-Matrix}} \mathbf{G}_{\text{parallel}}$ into \cref{eq:general_ls_mse}, the MSE is given by
\begin{align}
\text{MSE}_{\text{parallel}} &= \sigma_n^2 \text{Tr}\left( \left( \herm{\mathbf{A}_{\text{parallel}}} \mathbf{A}_{\text{parallel}} \right)^{-1} \right) \nonumber \\
&= \sigma_n^2 \text{Tr}\left( \mathbf{V} \right).
\label{eq:parallel_mse_h_expanded}
\end{align}

The S-Matrix activation scheme is derived from Hadamard matrix $\mathbf{W}_{\text{H-Matrix}}$ that is a scaled orthogonal matrix, satisfying
\begin{equation}
\herm{\mathbf{W}_{\text{H-Matrix}}} \mathbf{W}_{\text{H-Matrix}} = c \cdot \mathbf{I}_N,
\label{eq:s_matrix_orthogonality}
\end{equation}
where $c = N/4$ is a constant determined by the H-Matrix construction for $N$ being a power of 2. 
Thus, the parallel MSE under Hadamard matrix $\mathbf{W}_{\text{H-Matrix}}$ becomes
\begin{align}
\text{MSE}_{\text{parallel}}= \frac{\sigma_n^2}{c} \sum_{n=1}^{N} \frac{1}{|[\mathbf{g}_{\text{parallel}}]_n|^2}.
\label{eq:parallel_mse_h_simplified}
\end{align}

However, for passive PAs, the activation matrix $\mathbf{W}$ must be binary to be physically realizable, i.e., switching antennas ON/OFF. We therefore employ the S-Matrix proposed in the original system model. It is crucial to note that while derived from a Hadamard matrix, the S-Matrix is not orthogonal. Its Gram matrix, $\herm{\mathbf{W}_{\text{S-Matrix}}} \mathbf{W}_{\text{S-Matrix}}$, is a non-diagonal matrix. The non-zero off-diagonal elements introduce significant crosstalk between the parallel measurement paths.

Consequently, the Gram matrix of the observation matrix, $\herm{\mathbf{A}_{\text{parallel}}} \mathbf{A}_{\text{parallel}}$, does not simplify as in an ideal orthogonal case. The MSE must be evaluated using the general form:
\begin{equation}
\text{MSE}_{\text{parallel}} = \sigma_n^2 \text{Tr}( ( \herm{\mathbf{G}_{\text{parallel}}} \herm{\mathbf{W}_{\text{S-Matrix}}} \mathbf{W}_{\text{S-Matrix}} \mathbf{G}_{\text{parallel}} )^{-1} ),
\label{eq:parallel_mse_smatrix}
\end{equation}
where \begin{align}
\mathbf{W}^H \mathbf{W} &= \left(\frac{1}{2}(\mathbf{H} + \mathbf{J})\right)^H \left(\frac{1}{2}(\mathbf{H} + \mathbf{J})\right) \nonumber \\
&= \frac{1}{4}(\mathbf{H} + \mathbf{J})(\mathbf{H} + \mathbf{J}) \nonumber \\
&= \frac{1}{4}(\mathbf{H}\mathbf{H} + \mathbf{H}\mathbf{J} + \mathbf{J}\mathbf{H} + \mathbf{J}\mathbf{J}) \nonumber \\
&= \frac{1}{4} \left( N\mathbf{I}_N + \mathbf{H}\mathbf{J} + \mathbf{J}\mathbf{H} + N\mathbf{J} \right).
\end{align}
This result is a complex non-diagonal matrix. It is not a diagonal matrix—let alone a scalar multiple of the identity matrix. Thus, there is no constant \( c \) such that \( \mathbf{W}^H \mathbf{W} = c\mathbf{I}_N \). Unlike an orthogonal case which results in a simple sum of inverse squared norms, this expression is complex. The matrix inversion involves the full structure of both the in-waveguide channel matrix $\mathbf{G}_{\text{parallel}}$ and the crosstalk matrix $\herm{\mathbf{W}_{\text{S-Matrix}}} \mathbf{W}_{\text{S-Matrix}}$. The interaction between the crosstalk introduced by the S-Matrix and the ill-conditioning of $\mathbf{G}_{\text{parallel}}$ due to cumulative leakage can lead to severe noise amplification, significantly impacting the estimation performance.


\subsection{Measurement SNR Analysis}
\subsubsection{Measurement SNR for Serial Estimation}The received signal power of serial estimation in the $t$-th pilot slot, where only the $t$-th PA is active, is given by
\begin{equation}
P_{\text{serial},t} = |[\mathbf{b}_{\text{serial}}]_t|^2 = |[\mathbf{g}_{\text{serial}}]_t|^2 |[\mathbf{h}]_t|^2 = \gamma \alpha_{t}^2 e^{-2\varepsilon d_t} |[\mathbf{h}]_t|^2,
\label{eq:serial_signal_power}
\end{equation}
The corresponding measurement SNR is given by
\begin{equation}
\text{SNR}_{\text{serial},t} = \frac{P_{\text{serial},t}}{\sigma_n^2} = \frac{\gamma \alpha_{t}^2 e^{-2\varepsilon d_t} |[\mathbf{h}]_t|^2}{\sigma_n^2},
\label{eq:serial_snr}
\end{equation}

For parallel estimation, the $t$-th pilot slot activates $K = N/2$ PAs based on the property of S-Matrix schemes. Let $\mathcal{A}_t$ be the set of active PAs in slot $t$. The received signal power in slot $t$ is given by
\begin{equation}
P_{\text{parallel},t} = \left| \sum_{i \in \mathcal{A}_t} [\mathbf{b}_{\text{parallel}}]_i \right|^2,
\label{eq:parallel_signal_power_general}
\end{equation}
Assuming uncorrelated wireless channel paths ($\mathbb{E}[h_i \herm{h_j}] = 0$ for $i \neq j$), the expected power is the sum of individual powers:
\begin{align}
\mathbb{E}[P_{\text{parallel},t}] &= \sum_{i \in \mathcal{A}_t} |[\mathbf{b}_{\text{parallel}}]_i|^2 \nonumber \\
&= \sum_{i \in \mathcal{A}_t} \gamma \alpha_{i}^2 e^{-2\varepsilon d_i} |[\mathbf{h}]_i|^2 \left( \prod_{j=1}^{i-1} \beta_j^2 \right),
\label{eq:parallel_signal_power_expected}
\end{align}
The corresponding measurement SNR is given by
\begin{align}
\text{SNR}_{\text{parallel},t} &= \frac{\mathbb{E}[P_{\text{parallel},t}]}{\sigma_n^2} \nonumber \\
&= \frac{\gamma \alpha_{t}^2 }{\sigma_n^2} \sum_{i \in \mathcal{A}_t} e^{-2\varepsilon d_i} |[\mathbf{h}]_i|^2 \left( \prod_{j=1}^{i-1} \beta_j^2 \right),
\label{eq:parallel_snr}
\end{align}

Comparing \cref{eq:serial_snr} and \cref{eq:parallel_snr}, $\text{SNR}_{\text{parallel},t}$ is the sum of $N/2$ power terms, while $\text{SNR}_{\text{serial},t}$ is a single term. Even with cumulative leakage ($\prod_{j=1}^{i-1} \beta_j^2 < 1$ for $i>1$), the sum of $N/2$ array gain terms ensures $\mathbb{E}[P_{\text{parallel},t}] \gg P_{\text{serial},t}$ for functional systems. For parallel estimation to underperform, the cumulative leakage would need to reduce the sum of $N/2$ terms to less than one term, i.e., $\sum_{i \in \mathcal{A}_t} e^{-2\varepsilon d_i} \left( \prod_{j=1}^{i-1} \beta_j^2 \right) < e^{-2\varepsilon d_t}$ for all $t$. This is physically implausible, and hence parallel estimation achieves higher measurement SNR.

\subsection{Trade-off between Array Gain and Numerical Stability}
A superficial comparison of the MSE expressions is insufficient, as it obscures the critical role of the observation matrix conditioning—a key factor that dictates numerical stability and noise resilience. A deeper analysis reveals a fundamental trade-off between two competing effects: the array gain inherent to parallel PA activation and the numerical stability determined by the condition number of the observation matrix, which is strongly dependent on the signal pass-through coefficient $\beta$, i.e., a measure of inter-PA leakage.

\subsubsection{Serial Protocol}
For the serial protocol, the observation matrix is $\mathbf{A}_{\text{serial}} = \mathbf{G}_{\text{serial}}$, a diagonal matrix with elements $[\mathbf{g}_{\text{serial}}]_n = \sqrt{\gamma}\alpha_n e^{- \varepsilon d_n} e^{-j\frac{2\pi}{\lambda_g}d_n}$. The magnitude of these elements, $|[\mathbf{g}_{\text{serial}}]_n| = \sqrt{\gamma} e^{-\varepsilon d_n}$, decays polynomially with distance $d_n$ due to intrinsic waveguide attenuation $\varepsilon$, resulting in all diagonal elements being of a similar order of magnitude.
A diagonal matrix with elements of comparable magnitude is inherently well-conditioned. The condition number of $\mathbf{A}_{\text{serial}}$, i.e., ratio of maximum to minimum singular values is given by
\begin{equation}
\kappa(\mathbf{A}_{\text{serial}}) = \frac{\max_n |[\mathbf{g}_{\text{serial}}]_n|}{\min_n |[\mathbf{g}_{\text{serial}}]_n|} \approx e^{\varepsilon (d_{\text{max}} - d_{\text{min}})}.
\label{eq:serial_condition_number}
\end{equation}

The MSE for estimating the $n$-th wireless channel component $\hat{h}_n$ is inversely proportional to the measurement SNR of the $n$-th PA:
\begin{equation}
\text{MSE}(\hat{\mathbf{h}}_n)_{\text{serial}} \propto \frac{\sigma_n^2}{|[\mathbf{g}_{\text{serial}}]_n|^2} = \frac{\sigma_n^2}{\gamma \alpha_n e^{-2\varepsilon d_n}}.
\label{eq:serial_per_component_mse}
\end{equation}
While the serial protocol lacks array gain (each measurement uses only one PA), its well-conditioned observation matrix ensures estimation errors are gracefully bounded—noise is not amplified during matrix inversion, and errors are limited solely by the signal power of individual PAs.

\subsubsection{ Parallel Protocol}
The observation matrix of the parallel protocol is $\mathbf{A}_{\text{parallel}} = \mathbf{W}_{\text{S-Matrix}} \mathbf{G}_{\text{parallel}}$. The performance of the LS estimator depends critically on the condition number of $\mathbf{A}_{\text{parallel}}$, as the term $(\herm{\mathbf{A}_{\text{parallel}}} \mathbf{A}_{\text{parallel}})^{-1}$ in \cref{eq:parallel_mse_h_simplified} amplifies noise proportionally to $\kappa^2(\mathbf{A}_{\text{parallel}})$ (the square of the condition number).

The root of numerical instability lies in $\mathbf{G}_{\text{parallel}} = \text{diag}(\mathbf{g}_{\text{parallel}})$, whose elements decay exponentially with PA index $n$ due to cumulative leakage. From \cref{eq:parallel_transfer_complex}, assuming uniform leakage $\beta_i = \beta$ for all $i$, the magnitude of $\mathbf{G}_{\text{parallel}}$ elements is given by
\begin{equation}
|[\mathbf{g}_{\text{parallel}}]_n| = \sqrt{\gamma} \alpha_n e^{-\varepsilon d_n} \prod_{i=1}^{n-1} \beta_i \approx \sqrt{\gamma} \alpha_n e^{-\varepsilon d_n} \beta^{n-1}.
\label{eq:parallel_element_magnitude}
\end{equation}
For PAs far from the AP, i.e., a large $n$, the cumulative leakage term $\beta^{n-1}$ causes $|[\mathbf{g}_{\text{parallel}}]_n|$ to shrink drastically.

The condition number of $\mathbf{G}_{\text{parallel}}$ is dominated by this exponential decay:
\begin{equation}
\kappa(\mathbf{G}_{\text{parallel}}) = \frac{\max_n |[\mathbf{g}_{\text{parallel}}]_n|}{\min_n |[\mathbf{g}_{\text{parallel}}]_n|} \approx \frac{|[\mathbf{g}_{\text{parallel}}]_1|}{|[\mathbf{g}_{\text{parallel}}]_N|} \approx \frac{1}{\beta^{N-1}},
\label{eq:parallel_g_condition_number}
\end{equation}

While the S-Matrix $\mathbf{W}_{\text{S-Matrix}}$ is well-conditioned, premultiplying it with $\mathbf{G}_{\text{parallel}}$ cannot eliminate the extreme ill-conditioning inherited from $\mathbf{G}_{\text{parallel}}$. The condition number of $\mathbf{A}_{\text{parallel}}$ satisfies
\begin{equation}
\kappa(\mathbf{A}_{\text{parallel}}) \geq \frac{\kappa(\mathbf{G}_{\text{parallel}})}{\kappa(\mathbf{W}_{\text{S-Matrix}})}.
\label{eq:parallel_a_condition_number}
\end{equation}
Since $\kappa(\mathbf{W}_{\text{S-Matrix}}) = O(1)$, $\kappa(\mathbf{A}_{\text{parallel}})$ remains exponentially large for small $\beta$. This leads to catastrophic noise amplification when inverting $\herm{\mathbf{A}_{\text{parallel}}} \mathbf{A}_{\text{parallel}}$, completely overwhelming the array gain benefits of parallel PA activation.

\textbf{Remark 2:}
Based on the analysis above, the choice between serial and parallel estimation is not absolute—it depends on the physical leakage characteristics of the PASS hardware, quantified by $\beta$. This trade-off can be formally articulated as two regimes: 1) \textit{Low-leakage regime ($\beta \to 1$)}:  
    The cumulative leakage term $\beta^{n-1} \approx 1$ for all $n$, so $\mathbf{G}_{\text{parallel}}$ is well-conditioned ($\kappa(\mathbf{G}_{\text{parallel}}) \approx \kappa(\mathbf{G}_{\text{serial}})$). Noise amplification is minimal, and the array gain from coherently combining $N/2$ PAs per slot dominates. In this regime, parallel estimation achieves lower MSE than serial estimation. 2) \textit{High-leakage regime ($\beta \to 0$)}:  The cumulative leakage term causes $\mathbf{G}_{\text{parallel}}$ to become severely ill-conditioned. Noise amplification during matrix inversion completely nullifies array gain, leading to MSE values that are 10–100 dB worse than serial estimation. Here, the serial protocol—despite lacking array gain—provides a more stable and reliable estimate due to its inherently well-conditioned observation matrix.

\section{Performance Analysis of Uplink and Downlink Channel Estimation}
The channel estimation strategy for PANs is not monolithic, which is dictated by the fundamentally different physical constraints governing the downlink and uplink. In this section, we analyze them separately to reveal a profound operational asymmetry.

\subsection{Downlink Estimation: A Power-Budget-Limited Problem}
In the downlink, the AP possesses a finite total transmit power, $P_{\text{Total}}$, which must be allocated to create pilot signals. This power constraint is the defining factor for protocol performance.

\subsubsection{Serial Protocol}
The AP estimates the channel components sequentially in a serial protocol. During the $n$-th time slot, the sole objective is to probe the $n$-th channel component $\mathbf{h}_n$. Critically, the AP can dedicate its entire power budget $P_{\text{Total}}$ to this single measurement task. 

The pilot signal is launched into the waveguide and radiates exclusively from the $n$-th active PA (no other PAs are activated). Per the controllable radiation model in Section II.C.1, this path is free from cumulative leakage loss, i.e., only intrinsic waveguide attenuation $e^{-2\varepsilon d_n}$ and initial coupling loss $\gamma$ affect the signal. The received signal power at the UE for $\mathbf{h}_n$ is maximized, resulting in the highest possible measurement SNR for that channel component:
\begin{equation}
\text{SNR}_{\text{down, serial},n} = \frac{\gamma \alpha_n e^{-2\varepsilon d_n} P_{\text{Total}} |\mathbf{h}_n|^2}{\sigma_n^2}
\label{eq:downlink_serial_snr},
\end{equation}
where $\sigma_n^2$ is the noise variance of the user.

\subsubsection{Parallel Protocol}
In this case, the AP aims to estimate $G$ channel components simultaneously in a single time slot, e.g., $G=N/2$ for S-Matrix activation. The total power $P_{\text{Total}}$ must now be \textit{distributed} among these $K$ active PAs, i.e., each PA can receive at most $P_{\text{Total}}/G$ of the total power, assuming uniform power allocation and the optimal strategy for fairness and SNR balance.

This power division fundamentally reduces the strength of each individual pilot signal reaching the UE. Even though the S-Matrix protocol provides processing gain via orthogonal signal combination, it cannot recover from the initial power dilution. The measurement SNR for each channel component $\mathbf{h}_n$ is given by
\begin{align}
\text{SNR}_{\text{down, parallel},n} &= \frac{\gamma \alpha_n e^{-2\varepsilon d_n} (P_{\text{Total}}/G) |\mathbf{h}_n|^2}{\sigma_n^2} \nonumber \\
&= \frac{1}{G} \cdot \text{SNR}_{\text{down, serial},n}
\label{eq:downlink_parallel_snr}.
\end{align}

Downlink channel estimation is fundamentally a power allocation challenge. The serial protocol is unconditionally superior because its strategy of concentrating the entire power budget on one measurement at a time ensures the highest possible SNR for each channel component. The parallel protocol, by dividing power across multiple measurements, inherently operates in a lower SNR regime from which it cannot fully recover—processing gain cannot compensate for the fundamental power dilution.

\subsection{Uplink Estimation: An Energy-Collection vs. Stability Trade-off}
In the uplink, the UE transmits with a fixed power $P_{\text{UE}}$ (constrained by battery and regulatory limits), and its signal propagates to all $N$ PAs simultaneously. The problem shifts from power allocation (AP side) to energy collection (AP side), introducing a new trade-off between energy efficiency and numerical stability.

\subsubsection{Serial Protocol}
The serial protocol listens to only one PA at a time. During the $n$-th time slot, the AP activates only the $n$-th PA and discards the signal energy concurrently captured by the other $N-1$ PAs. This represents a significant loss of available signal energy, i.e., only a fraction of the transmitted power of UE is used for estimation.
However, the observation matrix $\mathbf{A}_{\text{serial}} = \mathbf{G}_{\text{serial}}$ of the serial protocol is inherently \textit{well-conditioned}. Its condition number $\kappa(\mathbf{A}_{\text{serial}}) \approx e^{\varepsilon (d_{\text{max}} - d_{\text{min}})}$ in Section IV.B.1 is small, ensuring the linear estimator is numerically robust. Noise amplification is negligible, and estimation errors are bounded by the single-PA SNR:
\begin{equation}
\text{SNR}_{\text{up, serial},n} = \frac{\gamma \alpha_n e^{-2\varepsilon d_n} P_{\text{UE}} |\mathbf{h}_n|^2}{\sigma_n^2}
\label{eq:uplink_serial_snr}.
\end{equation}

\subsubsection{Parallel Protocol}
The parallel protocol collects energy from $K$ PAs simultaneously, harnessing \textit{receive array gain}. The total signal energy available for estimation is the sum of energy from all active PAs, which can dramatically increase the measurement SNR. For uncorrelated signals, the combined SNR is approximately $G \cdot \text{SNR}_{\text{up, serial},n}$, promising a significant performance gain. However, this advantage is contingent upon the numerical stability of the parallel observation matrix $\mathbf{A}_{\text{parallel}} = \mathbf{W}_{\text{S-Matrix}} \mathbf{G}_{\text{parallel}}$. As rigorously analyzed in Section IV.B.2, the conditioning of $\mathbf{A}_{\text{parallel}}$ is dominated by $\mathbf{G}_{\text{parallel}}$, i.e., a diagonal matrix whose elements decay exponentially with PA index $n$ due to cumulative leakage:
\begin{equation}
|[\mathbf{G}_{\text{parallel}}]_n| \propto \beta^{n-1}
\label{eq:g_parallel_decay}.
\end{equation}
The condition number of $\mathbf{G}_{\text{parallel}}$ explodes exponentially with decreasing leakage coefficient $\beta$:
\begin{equation}
\kappa(\mathbf{G}_{\text{parallel}}) = \frac{\max_n |[\mathbf{G}_{\text{parallel}}]_n|}{\min_n |[\mathbf{G}_{\text{parallel}}]_n|} \approx \beta^{-(N-1)}.
\label{eq:g_parallel_condition}
\end{equation}

When $\mathbf{A}_{\text{parallel}}$ is ill-conditioned, the matrix inversion in the LS estimator amplifies noise proportionally to $\kappa^2(\mathbf{A}_{\text{parallel}})$. This catastrophic noise amplification can completely nullify the array gain benefit, leading to MSE values that are orders of magnitude worse than the serial protocol.

\section{Numerical Results}

In our simulations, we consider a typical indoor mmWave communication scenario, where the carrier frequency is set to 60 GHz and \( N = 16 \) PAs. For the practical in-waveguide channel model, we set the waveguide attenuation to \( \epsilon = 0.1 \, \text{nep/m} \). The bidirectional power splitting factor is set to \( \gamma = 0.5 \), indicating that half of the coupled power propagates towards the access point. For the downlink, the feed coupling efficiency from the access point to the waveguide is \( \eta = 0.9 \). To conduct a comprehensive analysis of practical parallel strategies, we define two distinct hardware-level power distribution models. The first is the proportional power radiation model, which simulates a passive and unoptimized hardware implementation. In this model, all PAs share identical physical coupling characteristics. Consequently, the signal power coupled by each PA is proportional to the remaining signal power present in the waveguide at its location, leading to an exponential decay in signal strength for PAs further from the AP due to the cumulative leakage effect. The second is the equal-power radiation model, which represents an optimized hardware design. This model assumes that the total energy captured can be losslessly collected and then redistributed equally among all $N$ antenna channels for processing by adjusting the coupling coefficients of each PA. This approach, while preserving energy conservation for a fair comparison, constructs a perfectly well-conditioned observation matrix with uniform signal power across all channels.
We evaluate the normalized MSE (NMSE) of channel estimation, which is defined as
\begin{equation}
\text{NMSE} = \mathbb{E}\left[ \frac{\|\hat{\mathbf{h}} - \mathbf{h}\|_2^2}{\|\mathbf{h}\|_2^2} \right].
\label{eq:nmse_definition}
\end{equation}

\begin{figure}[!t]
\centering
\includegraphics[width=3.0in]{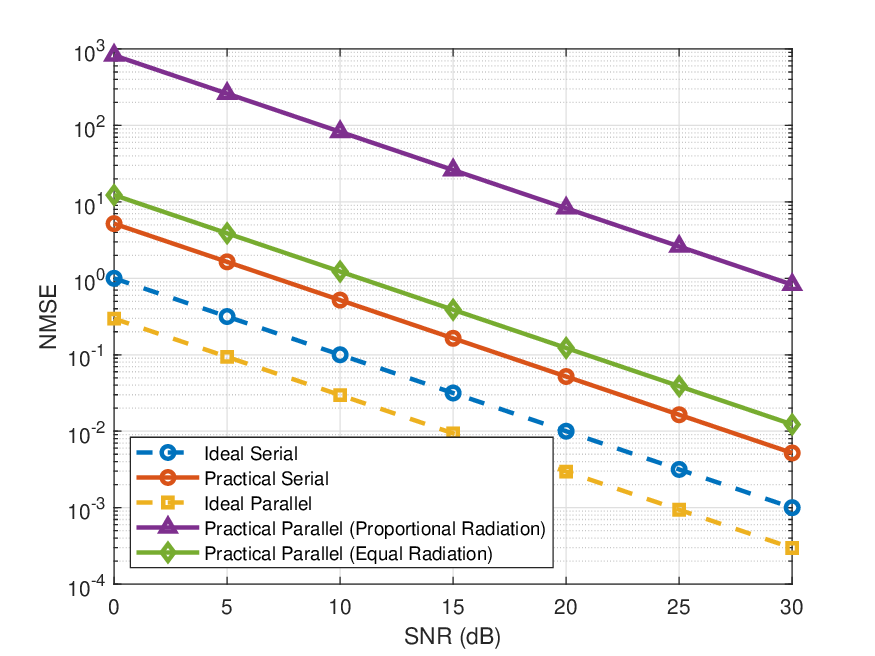} 
\caption{NMSE vs. SNR in uplink channel estimation.}
\label{fig:uplink_nmse}
\end{figure}

Fig. \ref{fig:uplink_nmse} illustrates the NMSE performance of uplink channel estimation for different operation protocols. The ideal parallel protocol significantly outperforms the ideal serial protocol, clearly demonstrating the $N$-fold array gain in a lossless environment. Among the practical schemes, the parallel protocol is inferior to the serial protocol as the severe crosstalk from the non-orthogonal S-Matrix compounds with the ill-conditioning from cumulative leakage. Compared to equal radiation model, the parallel protocol based on proportional radiation model has completely failed, which relies on ill-conditioned matrices, leading to catastrophic noise amplification and completely negating any potential array gain.

\begin{figure}[!t]
\centering
\includegraphics[width=3.0in]{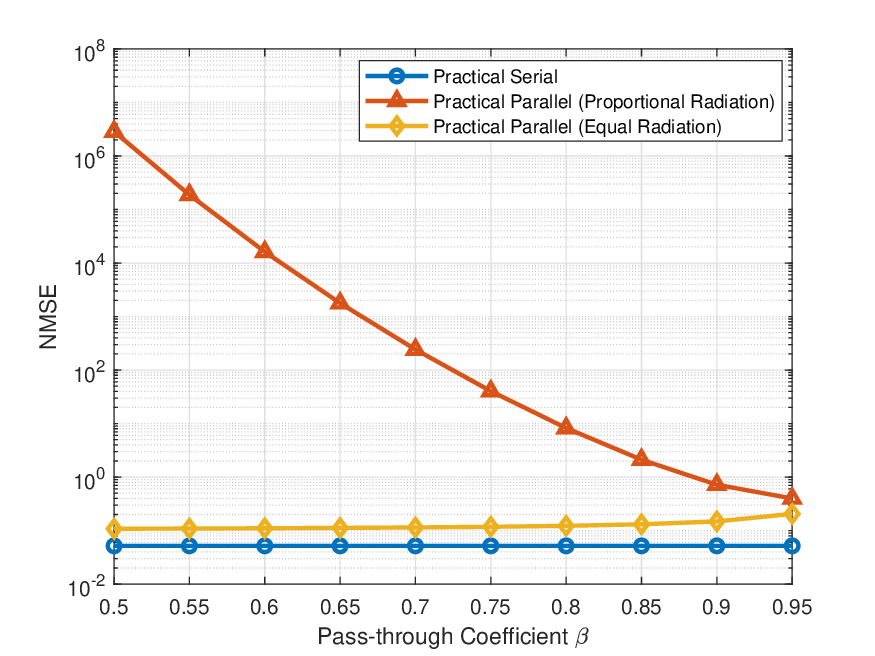} 
\caption{NMSE vs. pass-through coefficient $\beta$ in uplink channel estimation.}
\label{fig:nmse_vs_beta}
\end{figure}

Fig. \ref{fig:nmse_vs_beta} provides deeper insight into the uplink performance trade-off by presenting the NMSE as a function of the pass-through coefficient $\beta$. The performance of the practical serial protocol appears as a horizontal line, confirming its independence from inter-PA leakage and establishing it as a robust baseline. In stark contrast, the proportional radiation model shows extreme sensitivity to $\beta$. As $\beta$ decreases, the observation matrix becomes rapidly ill-conditioned, leading to catastrophic noise amplification and an exponential degradation in NMSE. The equal radiation model with numerical stability maintains the stable performance across nearly the entire range of $\beta$, while its performance slightly degrades as $\beta \to 1$ because the corresponding coupling coefficient $\alpha = \sqrt{1 - \beta^2}$ approaches zero, reducing the total energy coupled into the system.

\begin{figure}[!t]
\centering
\includegraphics[width=3.0in]{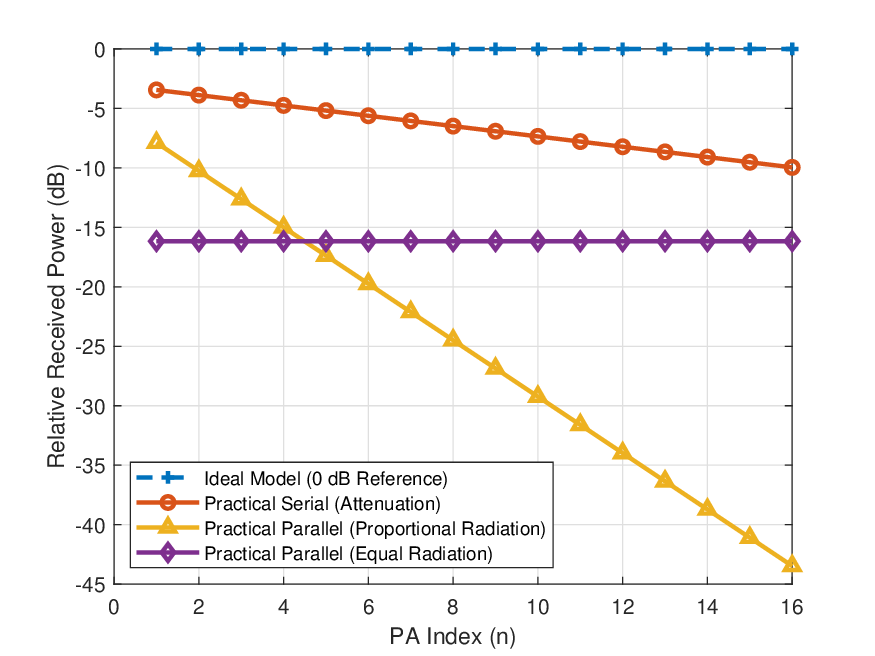} 
\caption{Received power decay across PAs for different operation protocols.}
\label{fig:power}
\end{figure}

Fig. \ref{fig:power} illustrates the physical process of signal power decay with respect to the PA index for different uplink models, providing a fundamental explanation for the performance disparities observed previously. The ideal model serves as a 0 dB reference, representing lossless transmission. The practical serial model exhibits a gentle, near-linear decay in the dB scale, accurately reflecting the physical characteristic that its signal is only affected by the intrinsic waveguide attenuation. In contrast, the power decay of the proportional radiation model is much more severe, where its steep slope reveals that the cumulative leakage effect is the primary cause of power loss for PAs far from the access point. Finally, the equal radiation model maintains a constant received power for all PAs. It is noteworthy that this constant power level is lower than the initial power of the other models, as it represents the result of averaging the total energy captured by the proportional radiation model across each PA, in adherence to the law of energy conservation. 

\begin{figure}[!t]
\centering
\includegraphics[width=3.0in]{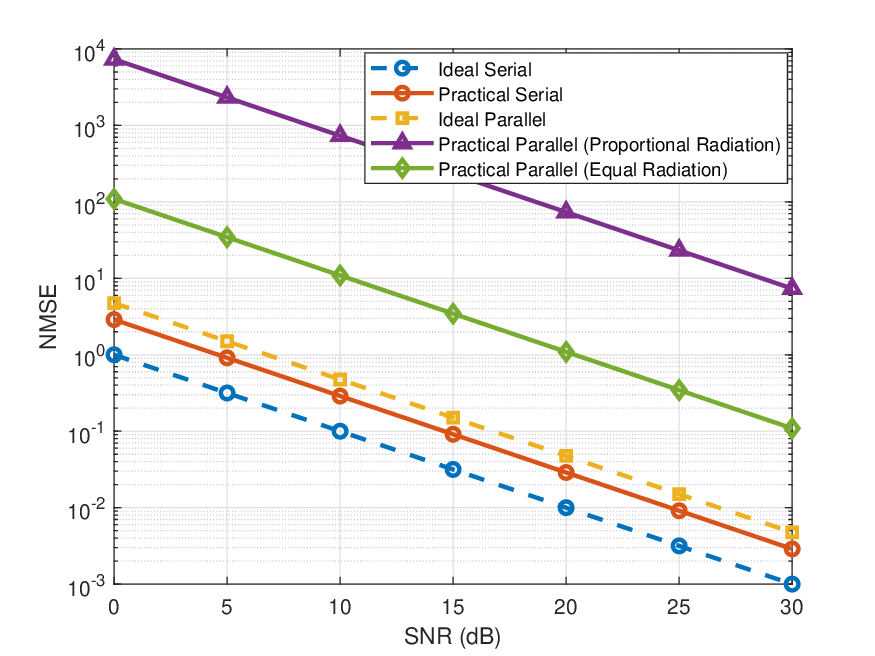} 
\caption{NMSE vs. SNR in downlink channel estimation.}
\label{fig:downlink}
\end{figure}

Fig. \ref{fig:downlink} shows the NMSE performance for the downlink channel estimation. The downlink estimation is fundamentally a power-budget-limited problem, where the serial protocol can concentrate the entire transmit power of the access point on a single PA measurement, thereby maximizing its SNR. Compared to uplink channel estimation, the downlink channel estimation can also avoid the bidirectional power splitting. In contrast, the parallel protocols must divide the total power among $N$ PAs, incurring an irrecoverable $1/N$ power allocation loss. Consequently, the serial protocol is the clear winner among all schemes in the downlink channel estimation. Note that the channel feedback overhead needs to be considered if the channel estimation is carried out at the user side.

\section{Conclusion}
In this paper, we investigated the channel estimation for PANs from the practical aspects. By analyzing a physically-grounded in-waveguide transmission model, we have provided the corresponding answers to our initial research questions.
\begin{itemize}

\item \textbf{A1:} \textit{Ideal vs. Practical Model:} Compared to ideal model, the practical uplink model introduces physical loss mechanisms, incorporating both the cumulative leakage along the propagation path and the initial bidirectional power split at the coupling point, which exacerbates the received SNR and numerical instability of the channel estimation process.

\item \textbf{A2:} \textit{Serial vs.  Parallel:} Compared to serial protocol in uplink channel estimation, the parallel protocol can utilize the array gain to improve channel estimation accuracy, while the performance of the parallel protocol is degraded by the structural crosstalk of non-orthogonal binary S-Matrix and numerical instability from cumulative leakage.

\item \textbf{A3:} \textit{Uplink vs. Downlink Estimation:} Compared to uplink channel estimation, the downlink estimation based on the serial protocol is more robust and achieves higher accuracy. This is because its performance is limited by a modest feed coupling loss $\eta$, whereas uplink estimation is penalized by a more significant bidirectional power splitting loss $\gamma$.

\end{itemize}

The channel estimation performance in PANs is not dictated by array gain alone but by a complex interplay with the numerical stability of the underlying physical model. Future hardware designs should prioritize minimizing leakage or implementing energy-balancing structures to fully unlock the potential of parallel processing in the uplink.

\bibliographystyle{IEEEtran}
\bibliography{IEEEabrv,refs.bib}
\end{document}